\documentclass[]{interact}

\usepackage{epstopdf}
\usepackage[caption=false]{subfig}
\usepackage{float}
\usepackage[numbers,sort&compress]{natbib}
\bibpunct[, ]{[}{]}{,}{n}{,}{,}
\makeatletter
\def\NAT@def@citea{\def\@citea{\NAT@separator}}
\makeatother

\theoremstyle{plain}

\theoremstyle{definition}

\usepackage{enumitem}
\usepackage{graphicx}

\theoremstyle{remark}

\begin{document}


\title{Anti-Jaynes-Cummings interaction of a two-level atom with squeezed light: A comparison with the Jaynes-Cummings interaction.}

\author{
\name{Christopher Mayero\textsuperscript{a,b} and Joseph Akeyo Omolo\textsuperscript{b}\thanks{Christopher Mayero. Email: cmayero@tmu.ac.ke} }
\affil{\textsuperscript{a}Tom Mboya University, Department of Physics, P.O. Box 199-40300, Homabay, Kenya.\\\textsuperscript{b}Maseno University, Department of Physics and Materials Science, Private Bag-40105, Maseno, Kenya.}
}
\maketitle

\begin{abstract}
We considered the anti-Jaynes-Cummings (AJC) interaction of a two-level atom in an initial ground state interacting with a field mode in an initial squeezed coherent state at arbitrary values of squeeze parameter \textit{r} and provided the Jaynes-Cummings (JC) interaction as a comparison. We analysed the degree of entanglement (DEM) measured by the von Neumann entropy and the nature of the field quantified by the Mandel Q parameter in relation to the atomic population inversion  during the AJC interaction and separately the corresponding JC interaction. We noted in our examples that at $r>1.4$, photon statistics evolved to super-Poissonian from sub-Poissonian during the respective AJC,\,JC interactions. Further, for high values of \textit{r}, the form of the time evolution of atomic population inversion depicted enhanced ringing revivals at the collapse region in comparison to the case of an initial coherent state. What is more, at higher values of \textit{r} the time evolution of DEM showed more rapid oscillations and recorded higher values, concurrently, an increase in the degree of mixedness. 
\end{abstract}


\section{Introduction}
\label{sec:introduction}
Interaction of an initial single mode of squeezed coherent light \cite{gerry2005introductory,Loudon1987Squeezed,Knight2004,zaheer1990advances} and a two-level atom in the JC interaction has received much attention leading to advanced studies of the interaction properties \cite{satyanarayana1989ringing,schleich1987oscillations,schleich1988area,teich1989squeezed}. 

In this present study we present a comparison, of presently, the AJC model \cite{omolo2021anti} and the well known JC model \cite{jaynes1963comparison} with respect to time evolution of DEM as measured by the von Neumann entropy and photon statistics quantified by the Mandel Q parameter during time evolution of atomic population inversion while varying the \textit{r} parameter at resonance and at constant corresponding AJC,\,JC field intensities. We relate variation in \textit{r} to the form of time evolution of atomic population inversion, DEM and photon statistics during the respective AJC,\,JC interactions.

This work is organised as follows; Sec.~\ref{sec:method} introduces the AJC,\,JC theoretical models and their time evolutions; in Sec.~\ref{sec:p_dit} comparison of the nature of photon statistics during the respective AJC,\,JC interactions is proffered; Sec.~\ref{sec:fent} discusses time evolution of DEM and atomic population inversion during the AJC process in relation to the corresponding JC process; and finally, Sec.~\ref{sec:conc} provides the conclusion.
\section{The models and their time evolution}
\label{sec:method}
The quantum Rabi model \cite{omolo2021anti} can be reorganised into a two-component (rotating and counter(anti)-rotating) form
\begin{subequations}
\begin{eqnarray}
\hat{H}_{Rabi}=\frac{1}{2}\left(\hat{H}+\hat{\overline{H}}\right)
\end{eqnarray} 
where the rotating part $\hat{H}$ is the normal order component, known as the JC Hamiltonian 
\begin{eqnarray}
\hat{H}=\hbar\Big[\omega\hat{N}+\delta\hat{s}_z+2\lambda(\hat{a}\hat{s}_++\hat{a}^\dagger\hat{s}_-)-\frac{1}{2}\omega\Big]\quad;\quad \hat{N}=\hat{a}^\dagger\hat{a}+\hat{s}_+\hat{s}_-\quad;\quad \delta=\omega_0-\omega\nonumber\\&&
\label{eq:RF1}
\end{eqnarray}
defining the coupling of a two-level system to the positive frequency field mode component. The standard conserved excitation number operator $\hat{N}$ in Eq.~\eqref{eq:RF1}, commutes with the Hamiltonian $\hat{H}$, i.e, $[\hat{N},\,\hat{H}]=0$. The counter(anti)-rotating part $\hat{\overline{H}}$ is the anti-normal order component known as the AJC Hamiltonian 
\begin{eqnarray}
\hat{\overline{H}}=\hbar\left[\omega\hat{\overline{N}}+\overline{\delta}\hat{s}_z+2\lambda(\hat{a}\hat{s}_-+\hat{a}^\dagger\hat{s}_+)-\frac{1}{2}\omega\right]\quad;\quad\hat{\overline{N}}=\hat{a}\hat{a}^\dagger+\hat{s}_-\hat{s}_+\quad;\quad\overline{\delta}=\omega_0+\omega\nonumber\\&&
\label{eq:CRF1}
\end{eqnarray}
that defines the coupling of a two-level system to the negative frequency field mode component \cite{born1999principles}. The conserved excitation number operator $\hat{\overline{N}}$ of the AJC interaction in Eq.~\eqref{eq:CRF1}, commutes with the Hamiltonian $\hat{\overline{H}}$, i.e, $[\hat{\overline{N}},\,\hat{\overline{H}}]=0$. Here, $\hat{s}_z,\,\hat{s}_+,\,\hat{s}_-$ are the atomic operators, $\hat{a}^\dagger,\,\hat{a}$ are the field operators, $\hbar$ the reduced Planks constant, $\omega_0$ the atomic state transition frequency and $\omega$ the quantised field mode angular frequency. 

In this work we are going to consider when a two-level atom is in an initial atomic ground state $|g\rangle$ during a resonant AJC ,\,JC interaction. To provide the desired comparison we define the sum frequency $\overline{\delta}=\omega_0+\omega$ in the AJC interaction in terms of frequency detuning $\delta=\omega_0-\omega$ in the JC interaction according to
\begin{eqnarray}
\overline{\delta}=\delta+2\omega
\end{eqnarray}
and so a resonant condition during the AJC interaction $\overline{\delta}=2\omega$, corresponds to  $\delta=0$ during the JC interaction.

%
%
%
%
\end{subequations}
 
Now, considering $|\psi_a\rangle_{t=0}$ as the generalised initial atomic state prepared in a superposition of excited $|e\rangle$ and ground state $|g\rangle$ prior to the JC,\,AJC interaction mechanism in the form
\begin{eqnarray}
|\psi_a\rangle_{t=0}=\sqrt{A}~|e\rangle+\sqrt{B}~|g\rangle
\end{eqnarray}
where the atom is in an excited state $|e\rangle$ with probability \textbf{A} and ground state $|g\rangle$ with a probability \textbf{B=1-A}, and $|\propto,\varsigma\rangle_{t=0}$   the state of the field initially prepared in a squeezed coherent state defined as \cite{gerry2005introductory} (where $\propto\equiv\alpha,\overline{\alpha}$ and $C_n\equiv S_n,\overline{S}_n$)

\begin{eqnarray}
|\propto,\varsigma\rangle_{t=0}&=&C_n|n\rangle \quad;\quad\nonumber\\
C_n&=&\frac{1}{\sqrt{\cosh(r)}}\exp\left[-\frac{1}{2}|\propto|^2-\frac{1}{2}\propto^{*2}e^{i\theta}\tanh(r)\right]\nonumber\\
&\times&\sum_{n=0}^\infty\frac{\left[\frac{1}{2}e^{i\theta}\tanh(r)\right]^{\frac{n}{2}}}{\sqrt{n!}}\times H_n\left[\left(\propto\cosh(r)+\propto^{*}e^{i\theta}\sinh(r)\right)\left(e^{i\theta}\sinh(2r)\right)^{-\frac{1}{2}}\right]\nonumber\\&&
\label{eq:sqc}
\end{eqnarray}
we easily define at \textbf{B=1,\,A=0} the JC and the AJC initial atom-field qubit state vectors $|\psi_{gn}\rangle_{JC},\,|\psi_{gn}\rangle_{AJC}$ obtained as direct product of atom, field quantum systems \cite{scully1997quantum} according to 
\begin{subequations}
\begin{eqnarray}
|\psi_{gn}\rangle_{AJC(JC)}=|\psi_g\rangle\otimes|\propto,\varsigma\rangle~,
\end{eqnarray}
to obtain
\begin{eqnarray}
|\psi_{gn}\rangle_{JC}&=&\sum_{n=0}^\infty S_n~|g,n\rangle\quad;\quad |\psi_{gn}\rangle_{AJC}=\sum_{n=0}^\infty\overline{S}_n~|g,n\rangle~;\nonumber\\
\langle\hat{a}^\dagger\hat{a}\rangle_{t=0}&=&|\alpha|^2+\sinh^2(r)\quad;\quad \langle\hat{a}\hat{a}^\dagger\rangle_{t=0}=\langle 1+\hat{a}^\dagger\hat{a}\rangle_{t=0}=|\overline{\alpha}|^2+\sinh^2(r)~.\nonumber\\&&
\label{eq:initial}
\end{eqnarray}
\end{subequations}

The probability of finding \textit{n} photons in the field is given by \cite{gerry2005introductory}
\begin{eqnarray}
P_n=\overline{P}_n&=&\vert\langle n\vert \propto,\varsigma\rangle\vert^2\nonumber\\
&=&\frac{\left[\frac{1}{2}\tanh(r)\right]^n}{n!\cosh(r)}\exp\left[-|\propto|^2-\frac{1}{2}\left(\propto^{*2}e^{i\theta}+\propto^2e^{-i\theta}\right)\tanh(r)\right]\nonumber\\
&\times&\Big\vert H_n\left[(\propto\cosh(r)+\propto^*e^{i\theta}\sinh(r))(e^{i\theta}\sinh(2r))^{-\frac{1}{2}}\right]\Big\vert^2~.
\label{eq:photonno}
\end{eqnarray}

Referring to Eq.~\eqref{eq:sqc}, $\varsigma=r\exp(i\theta)$ is the complex squeeze parameter, $r$ the squeeze parameter and $\alpha,\,\overline{\alpha}$ the JC,\,AJC coherent amplitude. Here, we consider an initial squeezed coherent state with $\theta=0$, and so $\varsigma=r$, $\alpha,\,\overline{\alpha}$ are  real. This implies that the generalised squeezed coherent state $|\propto,\varsigma\rangle$ is now mapped onto $|\propto,r\rangle$. 

The exact solution  $|\Psi_{gn}(t)\rangle,\,|\overline{\Psi}_{gn}(t)\rangle$ to the Schrödinger equation \cite{omolo2021anti} for the JC,\,AJC initial atom-field system in Eq.~\eqref{eq:initial} take the explicit forms ($t>0$)
\begin{subequations}
\begin{eqnarray}
|\Psi_{gn}(t)\rangle&=&e^{-\frac{i}{\hbar}\hat{H}t}|g,n\rangle=\sum_{n=0}^\infty\Big[S_n~e^{-i\omega nt}~\Big(\cos(R_{gn}t)\nonumber\\&&+ic_{gn}\sin(R_{gn}t)\Big)|g\rangle-ie^{-i\omega \left(n+1\right)t}~S_{n+1}~s_{gn+1}\sin(R_{gn+1})|e\rangle\Big]\otimes|n\rangle~;\nonumber\\
R_{gn}&=&\frac{\lambda}{2}\sqrt{4n+\beta^2}\quad;\quad c_{gn}=\frac{\beta}{\sqrt{4n+\beta^2}}\quad;\quad s_{gn}=2\sqrt{\frac{n}{4n+\beta^2}}\quad;\quad \beta=\frac{\delta}{\lambda}\nonumber\\&&
\label{eq:jcev}
\end{eqnarray}
and 
\begin{eqnarray}
|\overline{\Psi}_{gn}(t)\rangle&=&e^{-\frac{i}{\hbar}\hat{\overline{H}}t}~|g,n\rangle=\sum_{n=0}^\infty\Big[e^{-i\omega \left(n+1\right)t}~\overline{S}_n~\Big(\cos(\overline{R}_{gn}t)\nonumber\\&&+i\overline{c}_{gn}\sin(\overline{R}_{gn}t)\Big)|g\rangle-ie^{-i\omega nt}~\overline{S}_{n-1}~\overline{s}_{gn-1}\sin(\overline{R}_{gn-1})|e\rangle\Big]\otimes|n\rangle~;\nonumber\\
\overline{R}_{gn}&=&\frac{\lambda}{2}\sqrt{4n+4+(\beta+2\xi)^2}\quad;\quad\overline{c}_{gn}=\frac{(\beta+2\xi)}{\sqrt{4n+4+(\beta+2\xi)^2}}\nonumber\\
\overline{s}_{gn}&=&\sqrt{\frac{4(n+1)}{4n+4+(\beta+2\xi)^2}}\quad;\quad\overline{\delta}=\delta+2\omega\quad;\quad\xi=\frac{\omega}{\lambda}\quad;\quad\beta=\frac{\delta}{\lambda}~.\nonumber\\&&
\label{eq:ajcev}
\end{eqnarray}
\end{subequations}
The final forms of Eqs.~\eqref{eq:jcev} and~\eqref{eq:ajcev} have been arrived at through Schmidt decomposition \cite{gerry2005introductory} and so entanglement of the two interacting atom, field quantum systems is readily apparent.
\section{Photon statistics}
\label{sec:p_dit}
We now examine in this section the nature of photons during the respective JC,\,AJC interactions by application of the Mandel Q parameter \cite{mandel1979sub,teich1988photon,teich1989squeezed,mandel1995optical}. The Mandel Parameter  is fundamental in characterising the quantum statistical properties of a system. It can be calculated by knowing the photon-number distribution of a quantum state. In the Fock space $\mathbb{H}_f$ it takes the general form

\begin{eqnarray}
Q=\frac{\langle\left(\Delta{\hat{\eta}}\right)^2\rangle}{\langle{\hat{\eta}}\rangle}-1\quad;\quad \Delta{\hat{\eta}}=\sqrt{\langle{\hat{\eta}}^2\rangle-\langle{\hat{\eta}}\rangle^2}
\label{eq:themandel}
\end{eqnarray}
where $\langle\left(\Delta{\hat{\eta}}\right)^2\rangle$ is the photon number variance, $\langle{\hat{\eta}}\rangle$ is the mean photon number and $\hat{\eta}\equiv\hat{a}^\dagger\hat{a},\,\hat{a}\hat{a}^\dagger$ are the normal, anti-normal order operators of the number of particles (excitations). We take note that the sign of Mandel parameter determines the nature of deviation of excitation statistics from the Poisson statistics. More precisely, the Mandel parameter is positive (\texttt{Q>0}) when the statistic is super-Poissonian, zero (\texttt{Q=0}) when Poissonian and negative (\texttt{Q<0}) when sub-Poissonian with values ranging between \texttt{0} and \texttt{-1} during which the phenomenon of anti-bunching occurs \cite{teich1988photon} a clear manifestation of quantum effect.

The initial average photon number in the normal \cite{gerry2005introductory}, anti-normal order $|\alpha|^2+\sinh^2(r),\,|\overline{\alpha}|^2+\sinh^2(r)$ in the JC,\,AJC processes are defined in Eq.\eqref{eq:initial}. As time advances the average photon number in the JC, AJC processes take the forms ($t>0$)
\begin{subequations}
\begin{eqnarray}
\langle\hat{a}^\dagger\hat{a}\rangle_{\textit{t}}=tr\left[\hat{\rho}_f(t)~\hat{a}^\dagger\hat{a}\right]\quad;\quad\langle\hat{a}\hat{a}^\dagger\rangle_{\textit{t}}=tr\left[\hat{\overline{\rho}}_f(t)~\hat{a}\hat{a}^\dagger\right]=tr\left[\hat{\overline{\rho}}_f(t)\left(1+\hat{a}^\dagger\hat{a}\right)\right]
\label{eq:meanpn}
\end{eqnarray}
and it therefore follows that
\begin{eqnarray}
 \langle(\hat{a}^\dagger\hat{a})^2\rangle_{\textit{t}}=tr\left[\hat{\rho}_f(t)~(\hat{a}^\dagger\hat{a})^2\right]\quad;\quad\langle(\hat{a}\hat{a}^\dagger)^2\rangle_{\textit{t}}=tr\left[\hat{\overline{\rho}}_f(t)\left(1+\hat{a}^\dagger\hat{a}\right)^2\right]~.
 \label{eq:msqrd}
\end{eqnarray}
\end{subequations}

The time evolving reduced density operators of the field $\hat{\rho}_f^g(t),\,\hat{\overline{\rho}}_f^g(t)$ in the JC,\,AJC interaction determined from Eqs.~\eqref{eq:jcev},~\eqref{eq:ajcev} are easily obtained explicitly as
\begin{subequations}
\begin{eqnarray}
\hat{\rho}_f^g(t)&=&tr_a(|\Psi_{gn}(t)\rangle\langle\Psi_{gn}(t)|)=\Big[S_n^2\Big(\cos^2(R_{gn}t)+c_{gn}^2\sin^2(R_{gn}t)\Big)\nonumber\\&&+S_{n+1}^2~s_{gn+1}^2\sin^2(R_{gn+1}t)\Big]\otimes|n\rangle\langle n|
\label{eq:jcred}
\end{eqnarray}
and
\begin{eqnarray}
\hat{\overline{\rho}}_f^g(t)&=&tr_a(|\overline{\Psi}_{gn}(t)\rangle\langle\overline{\Psi}_{gn}(t)|)=\sum_{n=0}^\infty\Big[\overline{S}_n^2\Big(\cos^2(\overline{R}_{gn}t)+\overline{c}_{gn}^2\sin^2(\overline{R}_{gn}t)\Big)\nonumber\\&&+\overline{S}_{n-1}^2~\overline{s}_{gn-1}^2\sin^2(\overline{R}_{gn-1}t)\Big]\otimes|n\rangle\langle n|~.
\label{eq:ajcred}
\end{eqnarray}
\end{subequations}

\begin{figure}[H]
\centering
\subfloat[JC]{\label{fig:JCmand1}
\centering
\includegraphics[scale=0.75]{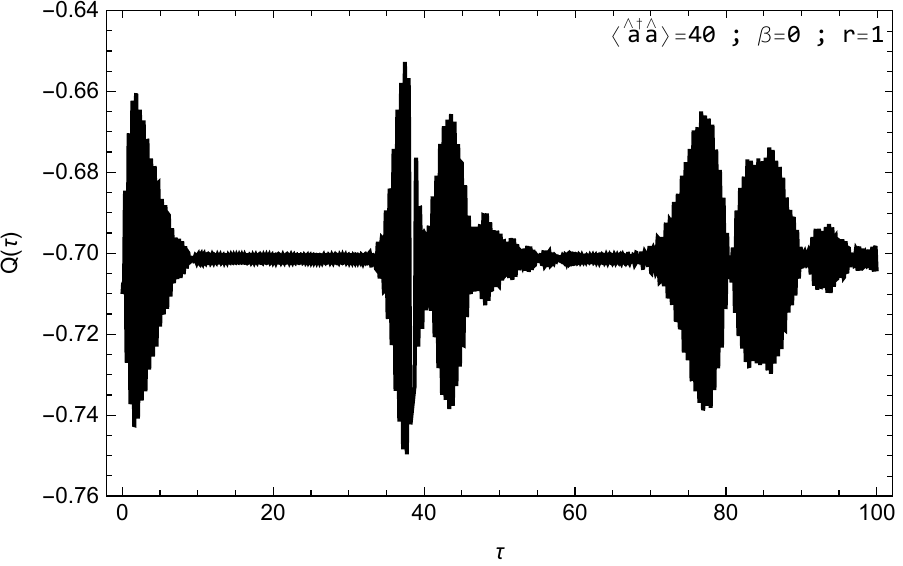}
}
\centering
\subfloat[AJC]{\label{fig:AJCmand1}
\centering
\includegraphics[scale=0.75]{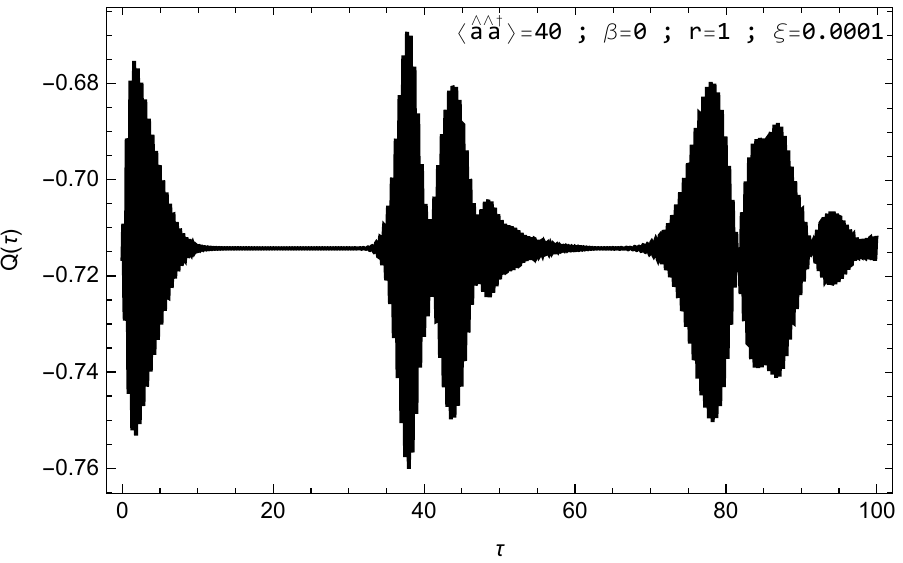}
}
\caption{Time evolution of Mandel parameter. Fig.~\eqref{fig:JCmand1},$Q(\tau)$ at $\beta=0,\,r=1$ and $|\alpha|^2+\sinh^2(r)=40$ in the JC interaction while Fig.~\eqref{fig:AJCmand1} is the corresponding time evolution of $Q(\tau)$ at  $2\xi\,;\,\beta=0,\,r=1$, $\xi=0.0001$ and $|\overline{\alpha}|^2+\sinh^2(r)=40$ in the AJC process}
\label{fig:2022mandel1}
\end{figure}

\begin{figure}[H]
\centering
\subfloat[JC]{\label{fig:JCmand2}
\centering
\includegraphics[scale=0.75]{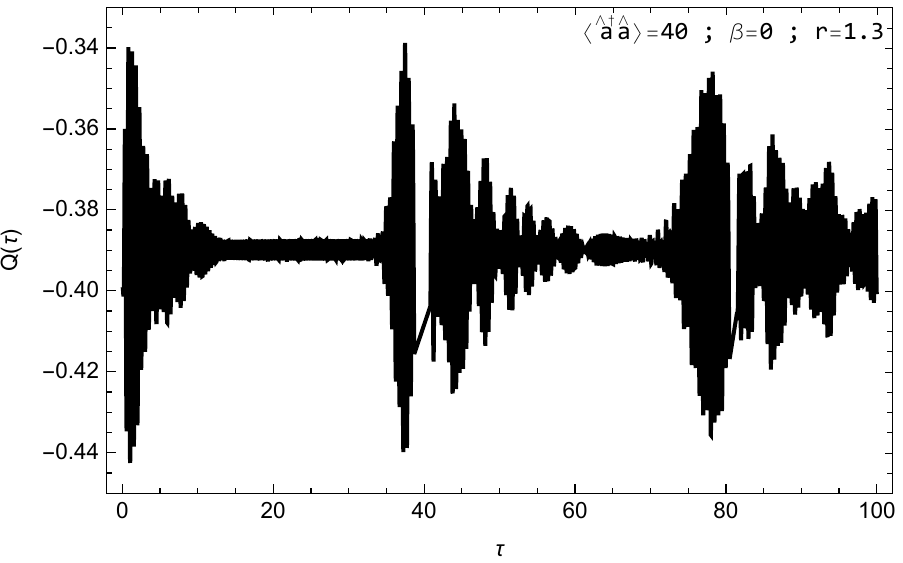}
}
\centering
\subfloat[AJC]{\label{fig:AJCmand2}
\centering
\includegraphics[scale=0.75]{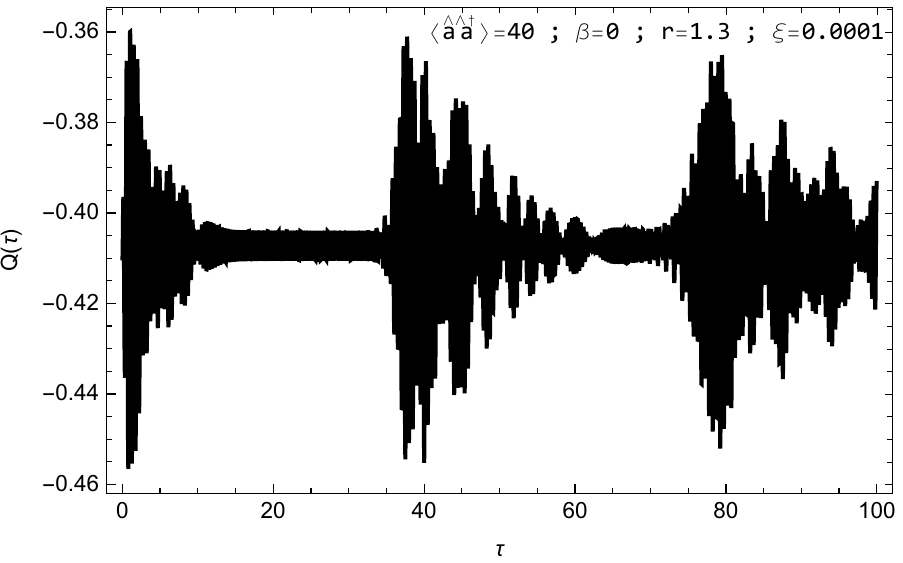}
}
\caption{Time evolution of Mandel parameter. Fig.~\eqref{fig:JCmand2},$Q(\tau)$ at $\beta=0,\,r=1.3$ and $|\alpha|^2+\sinh^2(r)=40$ in the JC interaction while Fig.~\eqref{fig:AJCmand2} is the corresponding time evolution of $Q(\tau)$ at  $2\xi\,;\,\beta=0,\,r=1.3$, $\xi=0.0001$ and $|\overline{\alpha}|^2+\sinh^2(r)=40$ in the AJC process}
\label{fig:2022mandel2}
\end{figure}

\begin{figure}[H]
\centering
\subfloat[JC]{\label{fig:JCmand3}
\centering
\includegraphics[scale=0.75]{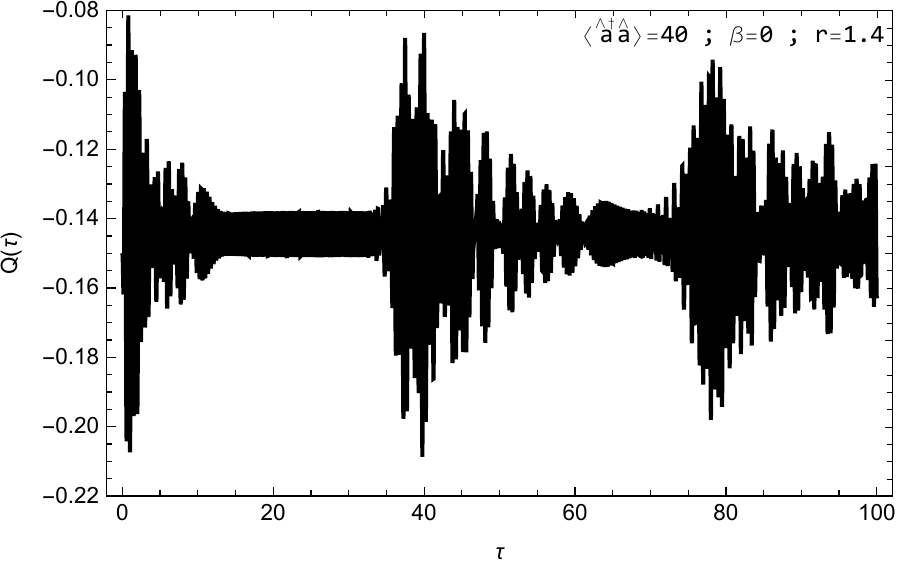}
}
\centering
\subfloat[AJC]{\label{fig:AJCmand3}
\centering
\includegraphics[scale=0.75]{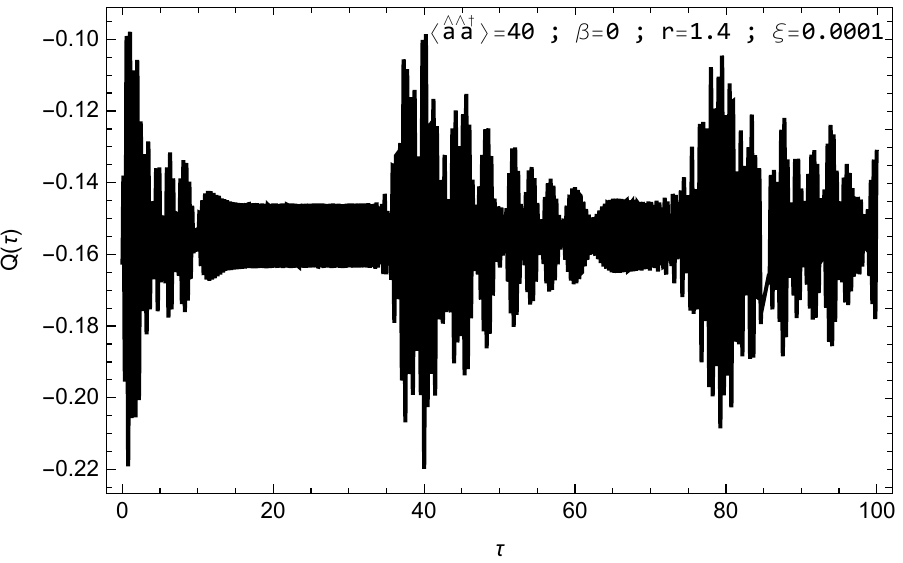}
}
\caption{Time evolution of Mandel parameter. Fig.~\eqref{fig:JCmand3},$Q(\tau)$ at $\beta=0,\,r=1.4$ and $|\alpha|^2+\sinh^2(r)=40$ in the JC interaction while Fig.~\eqref{fig:AJCmand3} is the corresponding time evolution of $Q(\tau)$ at  $2\xi\,;\,\beta=0,\,r=1.4$, $\xi=0.0001$ and $|\overline{\alpha}|^2+\sinh^2(r)=40$ in the AJC process}
\label{fig:2022mandel3}
\end{figure}

\begin{figure}[H]
\centering
\subfloat[JC]{\label{fig:JCmand4}
\centering
\includegraphics[scale=0.75]{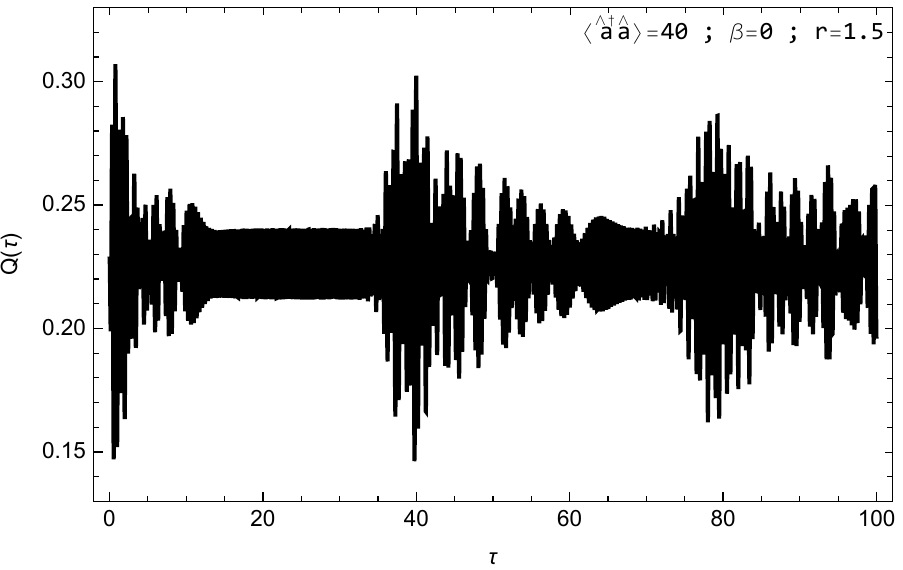}
}
\centering
\subfloat[AJC]{\label{fig:AJCmand4}
\centering
\includegraphics[scale=0.75]{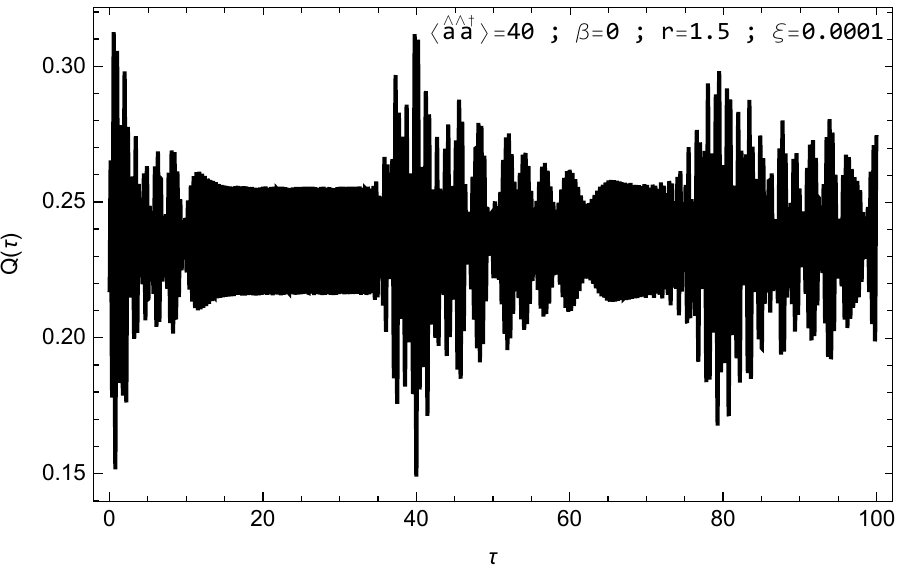}
}
\caption{Time evolution of Mandel parameter. Fig.~\eqref{fig:JCmand4},$Q(\tau)$ at $\beta=0,\,r=1.5$ and $|\alpha|^2+\sinh^2(r)=40$ in the JC interaction while Fig.~\eqref{fig:AJCmand4} is the corresponding time evolution of $Q(\tau)$ at  $2\xi\,;\,\beta=0,r=1.5$, $\xi=0.0001$ and $|\overline{\alpha}|^2+\sinh^2(r)=40$ in the AJC process}
\label{fig:2022mandel4}
\end{figure}

With the reduced field density operators in Eqs.~\eqref{eq:jcred},~\eqref{eq:ajcred}, interaction parameters, Rabi frequencies defined in Eqs.~\eqref{eq:jcev},~\eqref{eq:ajcev} and mean, mean square photon number defined in Eqs.~\eqref{eq:meanpn},~\eqref{eq:msqrd} we easily evaluate $Q(t)$ in Eq.~\eqref{eq:themandel} at resonance $\delta=0$ (JC), $\overline{\delta}=2\omega\,;\,\delta=0$ (AJC) and JC,\,AJC field intensity $|\alpha|^2+\sinh^2(r)=|\overline{\alpha}|^2+\sinh^2(r)=40$. We then plot time evolution of the Mandel Q parameter $Q(\tau)$ (where $\tau=\lambda t$ is the scaled time) for an initial atomic ground state $|g\rangle$ in an initial squeezed coherent state. Plots of the JC process at resonance $\delta=0\,;\,r=1,\,1.3,\,1.4,\,1.5$ and field intensity $|\alpha|^2+\sinh^2(r)=40$ are presented in Figs.~\ref{fig:JCmand1},~\ref{fig:JCmand2},~\ref{fig:JCmand3} and~\ref{fig:JCmand4} while the corresponding  AJC curves are provided in 
Figs.~\ref{fig:AJCmand1},~\ref{fig:AJCmand2},~\ref{fig:AJCmand3} and~\ref{fig:AJCmand4}.
 
From the plots in Figs.~\ref{fig:2022mandel1}~-~\ref{fig:2022mandel4} we see that;

\begin{enumerate}[label=\roman{*})]
\item {the photon statistics during the AJC interaction just like the JC interaction is dominantly sub-Poissonian at squeeze parameters $r=1$ to $r=1.4$ as presented in Figs.~\ref{fig:2022mandel1}~-~\ref{fig:2022mandel3}. The only exception is in Fig.~\ref{fig:2022mandel4} set at $r=1.5$ where the photon statistics evolves to a dominant super-Poissonian from sub-Poissonian photon statistics;} 
\item{in the AJC plots in Figs.~\ref{fig:AJCmand1},~\ref{fig:AJCmand2},~\ref{fig:AJCmand3} and~\ref{fig:AJCmand4}, the form of the time evolution of the Mandel parameter takes the same form as the corresponding JC cases in Figs.~\ref{fig:JCmand1},~\ref{fig:JCmand2},~\ref{fig:JCmand3} and~\ref{fig:AJCmand4}. However the evolutions of $Q(\tau)$ during the JC interactions in Figs.~\ref{fig:2022mandel1},~\ref{fig:2022mandel2} and~\ref{fig:2022mandel3} oscillate about mean positions different from the corresponding AJC $Q(\tau)$ evolutions. In the dominant super-Poissonian photon statistics in Fig.~\ref{fig:2022mandel4} during the AJC and separately JC interaction, $Q(\tau)$ oscillates about exact mean positions.}
\item{individual peaks in the AJC process record higher values of $Q(\tau)$ as plotted in Figs.~\ref{fig:AJCmand1},~\ref{fig:AJCmand2},~\ref{fig:AJCmand3} and~\ref{fig:AJCmand4}  than the corresponding JC peaks plotted in Figs.~\ref{fig:JCmand1},~\ref{fig:JCmand2},~\ref{fig:JCmand3} and~\ref{fig:AJCmand4}.}
\end{enumerate}
We conclude at this point that the interaction feature of the nature of photon statistics during the AJC process is similar to that realised during the JC interaction when an initial squeezed coherent field mode is considered.
\section{Evolution of atomic population inversion and von Neumann entropy}
\label{sec:fent}

To describe the evolution of the atom alone we introduce the reduced density matrices of the atom by tracing the JC,\,AJC density operators $\hat{\rho}_{gn},\,\hat{\overline{\rho}}_{gn}$ over the field states determined from Eqs.~\eqref{eq:jcev},~\eqref{eq:ajcev} according to
\begin{subequations}
\begin{eqnarray}
\hat{\rho}_a^g(t)=tr_f\left(|\Psi_{gn}(t)\rangle\langle\Psi_{gn}(t)|\right)\quad;\quad \hat{\overline{\rho}}_a^g(t)=tr_f\left(|\overline{\Psi}_{gn}(t)\rangle\langle\overline{\Psi}_{gn}(t)|\right)
\end{eqnarray}
taking explicit forms
\begin{eqnarray}
\hat{\rho}_a^g(t)&=&\sum_{n=0}^\infty\Big[S_n^2\Big(\cos^2(R_{gn}t)+c_{gn}^2\sin^2(R_{gn}t)\Big)|g\rangle\langle{g}|\nonumber\\&&
+i~S_n~S_{n+1}~s_{gn+1}e^{i\omega t}\sin(R_{gn+1}t)\Big(\cos(R_{gn}t)+ic_{gn}\sin(R_{gn}t)\Big)|g\rangle\langle e|\nonumber\\&&
-i~S_n~S_{n+1}~s_{gn+1}e^{-i\omega t}\sin(R_{gn+1}t)\Big(\cos(R_{gn}t)-ic_{gn}\sin(R_{gn}t)\Big)|e\rangle\langle g|\nonumber\\&&
+S_{n+1}^2~s_{gn+1}^2\sin^2(R_{gn+1}t)|e\rangle\langle e|\Big]\nonumber\\&&
\end{eqnarray}
and
\begin{eqnarray}
\hat{\overline{\rho}}_a^g(t)&=&\sum_{n=0}^\infty\Big[ \overline{S}_n^2\Big(\cos^2(\overline{R}_{gn}t)+\overline{c}_{gn}^2\sin^2(\overline{R}_{gn}t)\Big)|g\rangle\langle{g}|\nonumber\\&&
+i~\overline{S}_n~\overline{S}_{n-1}~\overline{s}_{gn-1}e^{-i\omega t}\sin(\overline{R}_{gn-1}t)\Big(\cos(\overline{R}_{gn}t)+i\overline{c}_{gn}\sin(\overline{R}_{gn}t)\Big)|g\rangle\langle e|\nonumber\\&&
-i~\overline{S}_n~\overline{S}_{n-1}~\overline{s}_{gn-1}e^{i\omega t}\sin(\overline{R}_{gn-1}t)\Big(\cos(\overline{R}_{gn}t)-i\overline{c}_{gn}\sin(\overline{R}_{gn}t)\Big)|e\rangle\langle g|\nonumber\\&&
+\overline{S}_{n-1}^2~\overline{s}_{gn-1}^2\sin^2(\overline{R}_{gn-1}t)|e\rangle\langle e|\Big]~.\nonumber\\&&
\end{eqnarray} 
\end{subequations}

We then define  the time evolving Bloch vector in the JC interaction $\vec{r}(t)=r_x(t)\hat{i}+r_y(t)\hat{j}+r_z(t)\hat{k}$ with components easily evaluated as $r_x(t)=tr\left(\hat{\sigma}_x\hat{\rho}_a^g(t)\right),\,r_y(t)=tr\left(\hat{\sigma}_y\hat{\rho}_a^g(t)\right),\,r_z(t)=tr\left(\hat{\sigma}_z\hat{\rho}_a^g(t)\right)$ reducing to explicit forms
\begin{subequations}
\begin{eqnarray}
r_x(t)&=&\sum_{n=0}^\infty\Big[S_n~S_{n+1}~\Big(-2s_{gn+1}\sin(R_{gn+1}t)\cos(R_{gn}t)\sin(\omega t)-2s_{gn+1}c_{gn}\nonumber\\&&\sin(R_{gn+1}t)\sin(R_{gn}t)\cos(\omega t)\Big)\Big]~;\nonumber\\
r_y(t)&=&\sum_{n=0}^\infty\Big[~S_n~S_{n+1}~\Big(2s_{gn+1}\sin(R_{gn+1}t)\cos(R_{gn}t)\cos(\omega t)-2s_{gn+1}c_{gn}\nonumber\\&&\sin(R_{gn+1}t)\sin(R_{gn}t)\sin(\omega t)\Big)\Big]~;\nonumber\\
r_z(t)&=&\sum_{n=0}^\infty \Big[S_{n+1}^2~s_{gn+1}^2\sin^2(R_{gn+1}t)-S_n^2~\Big(\cos^2(R_{gn}t)+c_{gn}^2\sin^2(R_{gn}t)\Big)\Big]~.\nonumber\\&&
\label{eq:jccomp}
\end{eqnarray}
Similarly, the time evolving Bloch vector in the AJC process $\vec{\overline{r}}(t)=\overline{r}_x(t)\hat{i}+\overline{r}_y(t)\hat{j}+\overline{r}_z(t)\hat{k}$ with components obtained as $\overline{r}_x(t)=tr\left(\hat{\sigma}_x\hat{\overline{\rho}}_a^g(t)\right),\,\overline{r}_y(t)=tr\left(\hat{\sigma}_y\hat{\overline{\rho}}_a^g(t)\right),\,\overline{r}_z(t)=tr\left(\hat{\sigma}_z\hat{\overline{\rho}}_a^g(t)\right)$ taking respective explicit forms  
\begin{eqnarray}
\overline{r}_x(t)&=&\sum_{n=0}^\infty \Big[\overline{S}_n~\overline{S}_{n-1}~\Big(2\overline{s}_{gn-1}\sin(\overline{R}_{gn-1}t)\cos(\overline{R}_{gn}t)\sin(\omega t)-2\overline{s}_{gn-1}\overline{c}_{gn}\nonumber\\&&\sin(\overline{R}_{gn-1}t)\sin(\overline{R}_{gn}t)\cos(\omega t)\Big)\Big]~;\nonumber\\
\overline{r}_y(t)&=&\sum_{n=0}^\infty \Big[\overline{S}_n~\overline{S}_{n-1}\Big(2\overline{s}_{gn-1}\sin(\overline{R}_{gn-1}t)\cos(\overline{R}_{gn}t)\cos(\omega t)+2\overline{s}_{gn-1}\overline{c}_{gn}\nonumber\\&&\sin(\overline{R}_{gn-1}t)\sin(\overline{R}_{gn}t)\sin(\omega t)\Big)\Big]~;\nonumber\\
\overline{r}_z(t)&=&\sum_{n=0}^\infty \Big[\overline{S}_{n-1}^2~\overline{s}_{gn-1}^2\sin^2(\overline{R}_{gn-1}t)-\overline{S}_n^2~\Big(\cos^2(\overline{R}_{gn}t)+\overline{c}_{gn}^2\sin^2(\overline{R}_{gn}t)\Big)\Big]~.\nonumber\\&&
\label{eq:ajccomp}
\end{eqnarray}
\end{subequations}
We use the Bloch vector components in Eqs.~\eqref{eq:jccomp} and~\eqref{eq:ajccomp} to evaluate time evolution of atomic population inversion $W(t)$ and the time evolution of the von Neumann entropy $S_a(t)$ (as a measure of DEM).

The atomic population inversion  $W(t)$ \cite{scully1997quantum} is defined as the difference between the excited and ground state probabilities 
\begin{eqnarray}
W(t)=tr\left(\hat{\sigma}_z\hat{\rho}_a(t)\right)
\end{eqnarray}
which is of the exact form as the \textit{z}-component $r_z(t),\,\overline{r}_z(t)$ of the time evolving JC,\,AJC Bloch vectors  in Eqs.~\eqref{eq:jccomp},~\eqref{eq:ajccomp}.

\begin{figure}[H]
\centering
\subfloat[JC]{\label{fig:JCpty}
\centering
\includegraphics[scale=0.75]{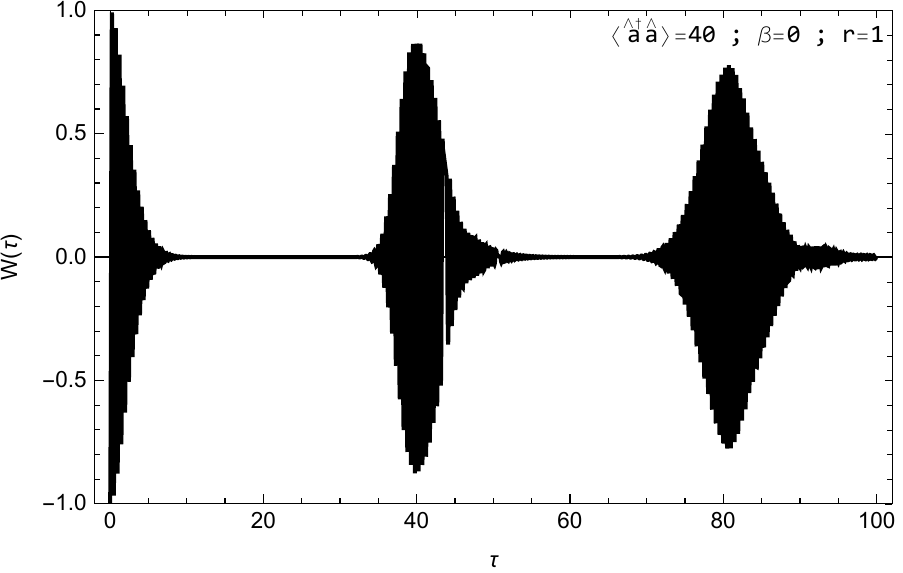}
}
\centering
\subfloat[AJC]{\label{fig:AJCpty}
\centering
\includegraphics[scale=0.75]{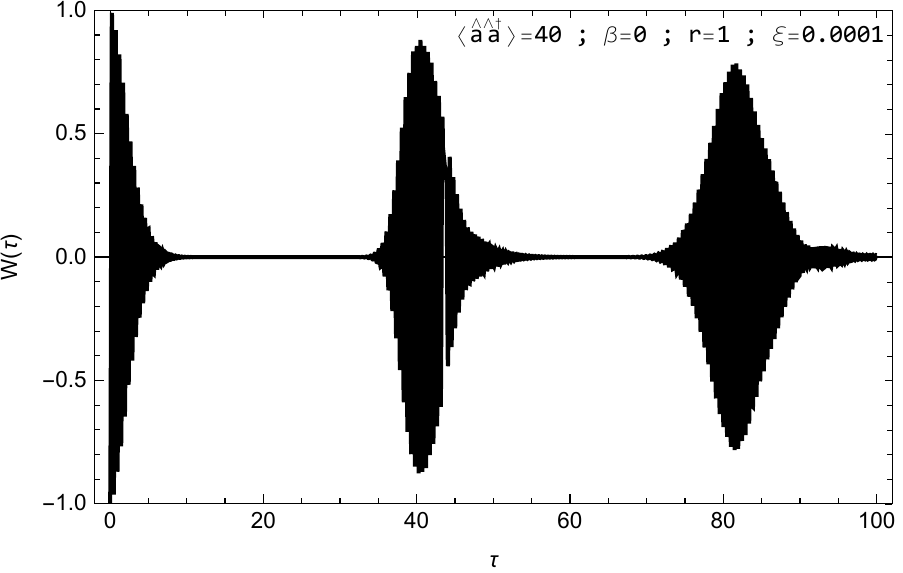}
}
\caption{Time evolution of atomic population inversion. Fig.~\eqref{fig:JCpty} $W(\tau)$ at $\beta=0,\,r=1$ and $|\alpha|^2+\sinh^2(r)=40$ in the JC interaction while Fig.~\eqref{fig:AJCpty} is the corresponding time evolution of $W(\tau)$ at  $2\xi\,;\,\beta=0$,\,$r=1$, $\xi=0.0001$ and $|\overline{\alpha}|^2+\sinh^2(r)=40$ in the AJC process}
\label{fig:2022pty}
\end{figure}

\begin{figure}[H]
\centering
\subfloat[JC]{\label{fig:mJCpty}
\centering
\includegraphics[scale=0.75]{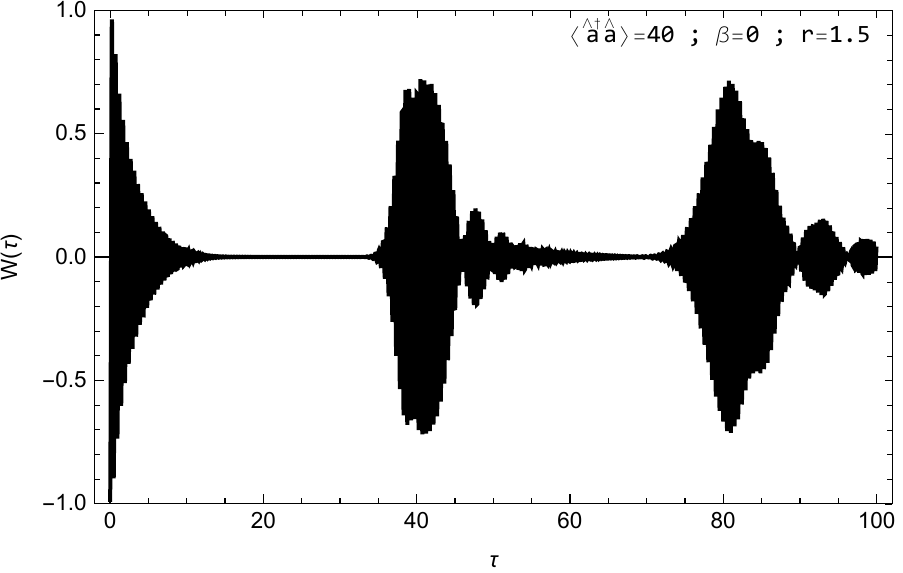}
}
\centering
\subfloat[AJC]{\label{fig:mAJCpty}
\centering
\includegraphics[scale=0.75]{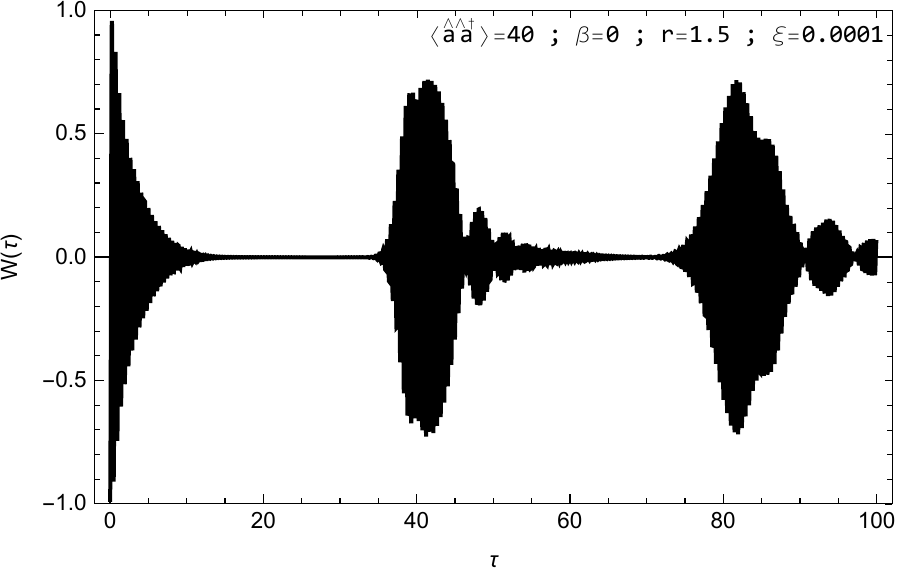}
}
\caption{Time evolution of atomic population inversion. Fig.~\eqref{fig:JCpty}, $W(\tau)$ at $\beta=0,\,r=1.5$ and $|\alpha|^2+\sinh^2(r)=40$ in the JC interaction while Fig.~\eqref{fig:AJCpty} is the corresponding time evolution of $W(\tau)$ at  $2\xi\,;\,\beta=0,\,r=1.5$, $\xi=0.0001$ and $|\overline{\alpha}|^2+\sinh^2(r)=40$ in the AJC process}
\label{fig:m2022pty}
\end{figure}

Using the definitions of $r_z(t),\,\overline{r}_z(t)$ in Eqs.~\eqref{eq:jccomp},~\eqref{eq:ajccomp}, we plot $W(t)$ in Figs.~\ref{fig:JCpty},~\ref{fig:mJCpty} at $\delta=0,r=1,\,1.5,\,|\alpha|^2+\sinh^2(r)=40$ during the JC interaction and in Figs.~\ref{fig:AJCpty},~\ref{fig:mAJCpty} at $\overline{\delta}=2\omega\,;\,\delta=0,r=1,\,1.5,|\overline{\alpha}|^2+\sinh^2(r)=40$ during the AJC interaction.

\begin{figure}[H]
\centering
\subfloat[JC]{\label{fig:koJCpty}
\centering
\includegraphics[scale=0.75]{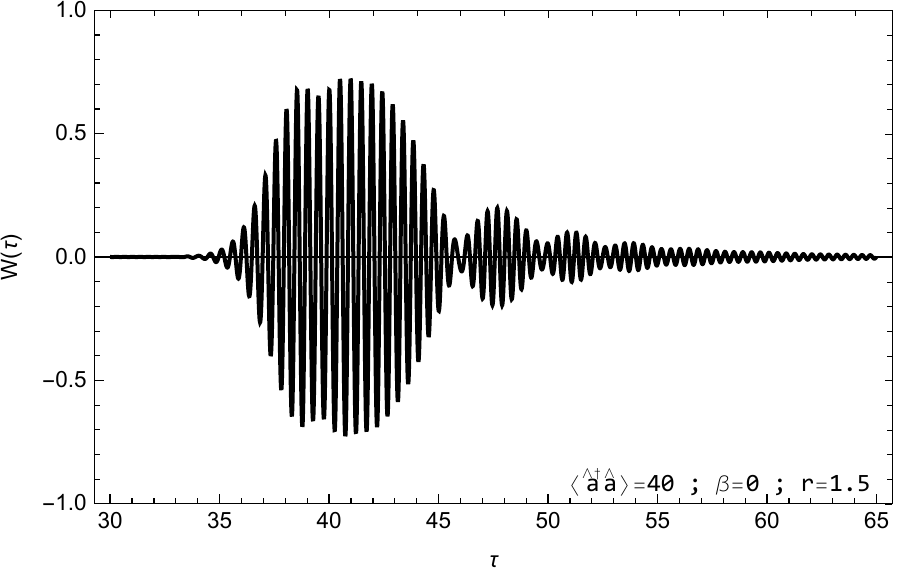}
}
\centering
\subfloat[AJC]{\label{fig:koAJCpty}
\centering
\includegraphics[scale=0.75]{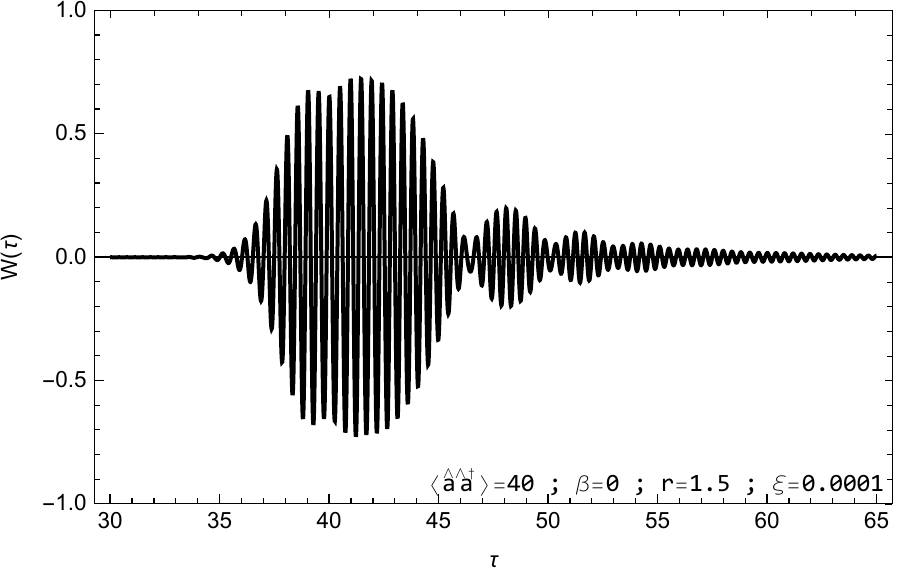}
}
\caption{Time evolution of atomic population inversion (Ringing revivals). Fig.~\eqref{fig:koJCpty}, $W(\tau)$ at $\beta=0,\,r=1.5$ and $|\alpha|^2+\sinh^2(r)=40$ in the JC interaction while Fig.~\eqref{fig:koAJCpty} is the corresponding time evolution of $W(\tau)$ at  $2\xi\,;\,\beta=0,\,r=1.5$, $\xi=0.0001$ and $|\overline{\alpha}|^2+\sinh^2(r)=40$ in the AJC process}
\label{fig:pp2022pty}
\end{figure}

\begin{figure}[H]
\centering
\subfloat[AJC and JC]{\label{fig:pnone}
\centering
\includegraphics[scale=0.75]{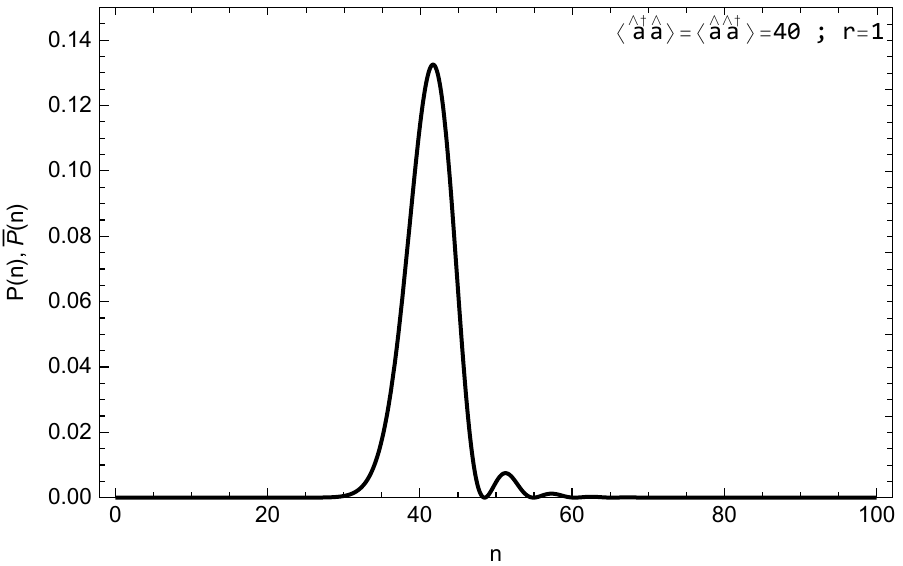}
}
\centering
\subfloat[AJC and JC]{\label{fig:pntwo}
\centering
\includegraphics[scale=0.75]{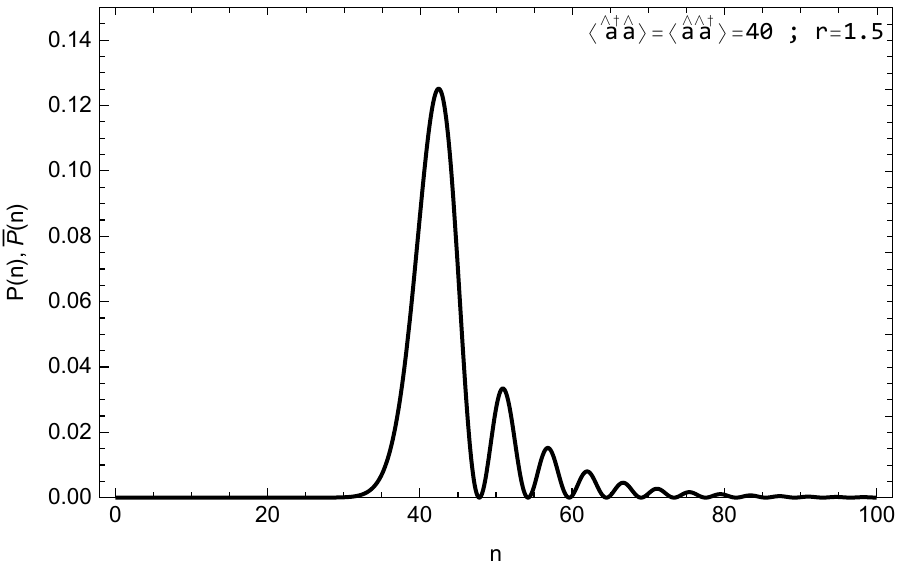}
}
\caption{JC,\,AJC photon number distribution $P_n,\,\overline{P}_n$. Fig.~\eqref{fig:pnone}, $P(n),\,\overline{P}(n)$ at $r=1$ and $|\propto|^2+\sinh^2(r)=40$ while Fig.~\eqref{fig:pntwo}  $P(n),\,\overline{P}(n)$ at  $r=1.5,\,|\propto|^2+\sinh^2(r)=40$}
\label{fig:2022pnno}
\end{figure}

We proceed to plot the photon number distribution $P_n=\overline{P}_n$ defined in Eq.~\eqref{eq:photonno} at $r=1,\,1.5,\,|\propto|^2+\sinh^2(r)=40$ in Figs.~\ref{fig:pnone},~\ref{fig:pntwo}.

From the results in Figs.~\ref{fig:2022pty}~-~\ref{fig:2022pnno} we see: 
\begin{enumerate}[label=\roman{*})]
\item that the time evolution of atomic population inversion during the AJC,\,JC processes display exact forms;
\item that the oscillations at $r=1.5$ in Figs.~\ref{fig:mJCpty},~\ref{fig:mAJCpty} are more irregular at the collapse region than when $r=1$ in Figs.~\ref{fig:JCpty},~\ref{fig:AJCpty}, commonly referred to as ringing revivals (see Fig.~\ref{fig:pp2022pty}) in agreement with \cite{satyanarayana1989ringing,moya1992interaction}, i.e., the collapse region is modulated or displays ringing different from the well known collapse region obtained when an initial coherent field is considered \cite{gerry2005introductory}. As explained in detail in \cite{satyanarayana1989ringing,moya1992interaction}, the ringing is due to interference of the additional peaks (see Fig.~\ref{fig:2022pnno}) in the JC photon number distribution $P_n=\vert\langle n\vert\alpha,r\rangle\vert^2$ and separately AJC  photon number distribution $\overline{P}_n=\vert\langle n\vert\overline{\alpha},r\rangle\vert^2$ because the revivals produced by different peaks of $P_n (JC),\,\overline{P}_n (AJC)$ have different local mean photon numbers. In the process revivals due to individual peaks overlap but the effect of the resulting interference is to sharpen the ringing structure other than washing it away, i.e, the addition of each local peak in $P_n,\,\overline{P}_n$ adds an echo in $W(\tau)$ and the successive echoes brings further interference, which sharpens the echoes at earlier times and;

\item{in Fig.~\ref{fig:2022pty} sharpness of the revival regions during atomic population inversion which occur when the field is sub-Poissonian (see Fig.~\ref{fig:2022mandel1} at $r=1$) accordant with \cite{moya1992interaction}, in comparison to the less pronounced and blunt peaks in Fig.~\ref{fig:m2022pty} at $r=1.5$. We noted in our example in Fig.~\ref{fig:2022mandel4} plotted at $r=1.5$ during the AJC,\,JC processes, that the field is super-Poissonian.} 
\end{enumerate} 

\begin{figure}[H]
\centering
\subfloat[JC]{\label{fig:JCfent1}
\centering
\includegraphics[scale=0.75]{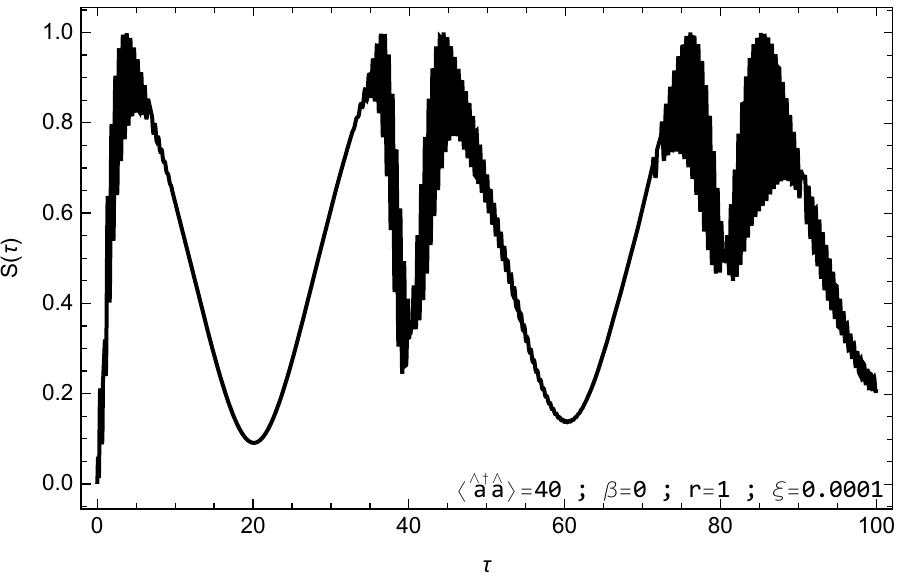}
}
\centering
\subfloat[AJC]{\label{fig:AJCfent1}
\centering
\includegraphics[scale=0.75]{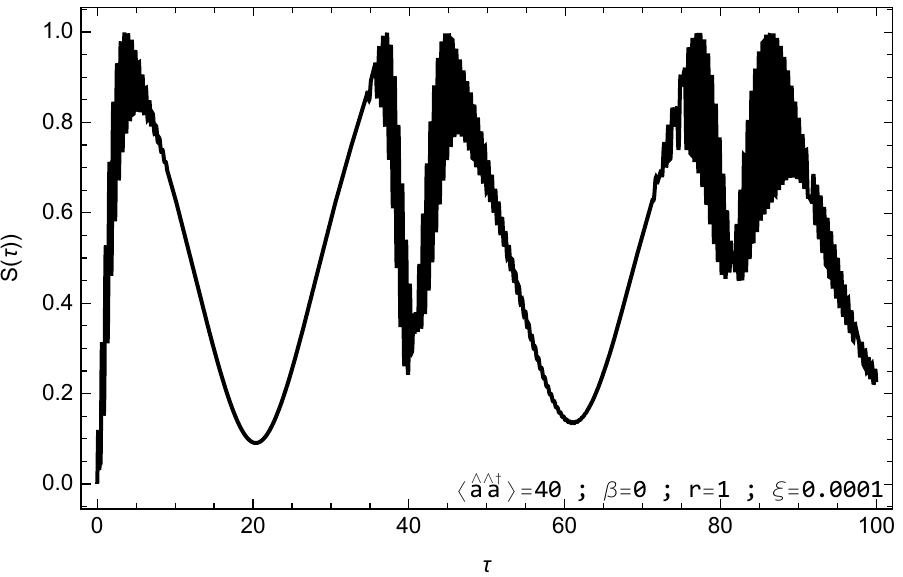}
}
\caption{Time evolution of atomic entropy. Fig.~\eqref{fig:JCfent1}, $S_a(\tau)$ at $\beta=0,\,r=1$ and $|\alpha|^2+\sinh^2(r)=40$ in the JC interaction while Fig.~\eqref{fig:AJCfent1} is the corresponding time evolution of $S_a(\tau)$ at  $2\xi\,;\,\beta=0,r=1$, $\xi=0.0001$ and $|\overline{\alpha}|^2+\sinh^2(r)=40$ in the AJC process}
\label{fig:2022fent1}
\end{figure}

\begin{figure}[H]
\centering
\subfloat[JC]{\label{fig:JCfent2}
\centering
\includegraphics[scale=0.75]{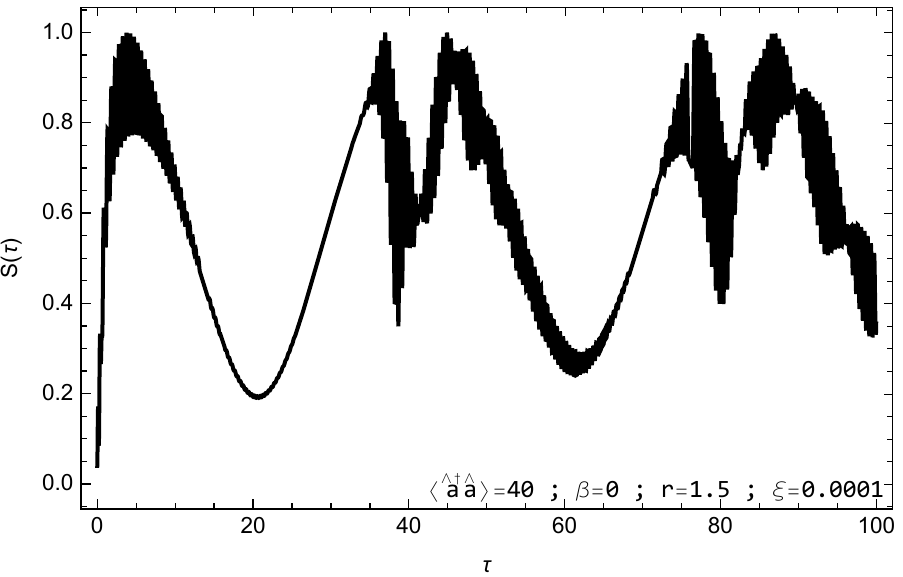}
}
\centering
\subfloat[AJC]{\label{fig:AJCfent2}
\centering
\includegraphics[scale=0.75]{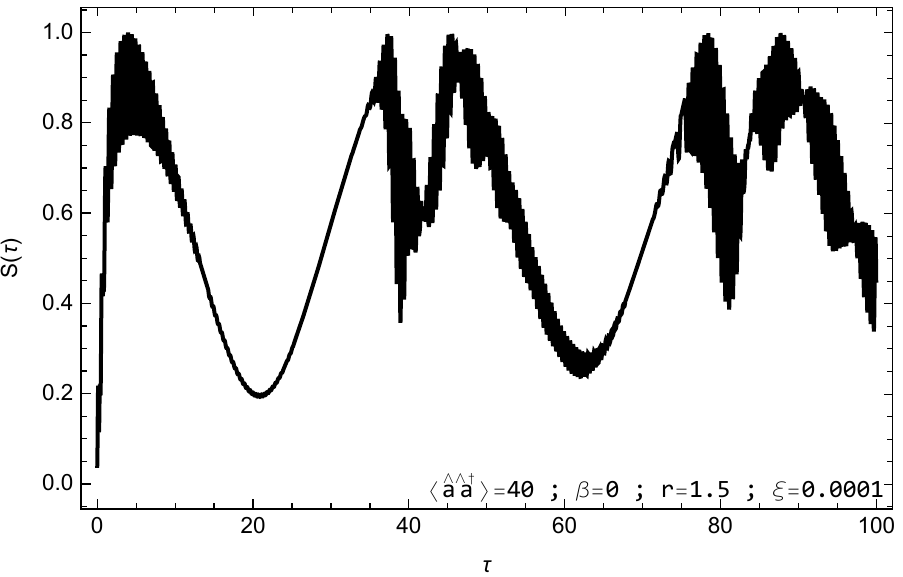}
}
\caption{Time evolution of atomic entropy. Fig.~\eqref{fig:JCfent2}, $S_a(\tau)$ at $\beta=0,\,r=1.5$ and $|\alpha|^2+\sinh^2(r)=40$ in the JC interaction while Fig.~\eqref{fig:AJCfent2} is the corresponding time evolution of $S_a(\tau)$ at  $2\xi\,;\,\beta=0,r=1.5$, $\xi=0.0001$ and $|\overline{\alpha}|^2+\sinh^2(r)=40$ in the AJC process}
\label{fig:2022fent2}
\end{figure}

Further, in order to discuss the collapses and revival phenomenon in relation to degree of entanglement we introduce the von Neumann entropy $S_a(t)$ defined in terms of the time evolving Bloch vector $\vec{r}(t)$ in the general form \cite{gerry2005introductory}
\begin{subequations}
\begin{eqnarray}
S_a(t)=-\eta_1\log_2\eta_1-\eta_2\log_2\eta_2
\end{eqnarray} 
where here
\begin{eqnarray}
\eta_1=\frac{1}{2}\left[1-\vert \vec{r}(t)\vert\right]\quad;\quad \eta_2=\frac{1}{2}\left[1+\vert\vec{r}(t)\vert\right]~.
\end{eqnarray}
\end{subequations}
The JC process atomic entropy plots at resonance $\delta=0,r=1,\,1.5,$ and field intensity $|\alpha|^2+\sinh^2(r)=40$ are shown in Figs.~\ref{fig:JCfent1} and~\ref{fig:JCfent2}  while the corresponding AJC curves are in Figs.~\ref{fig:AJCfent1} and~\ref{fig:AJCfent2}.

Based on the results in Figs.~\ref{fig:2022fent1} and~\ref{fig:2022fent2} we see;
\begin{enumerate}[label=\roman{*})]
\item{in Fig.~\ref{fig:2022fent1} during the JC,\,AJC interactions, the value of $S_a(\tau)$ at the revival time \cite{eberly1980}   $\tau_R=2\pi\sqrt{\vert\alpha\vert^2+\sinh^2(r)}\simeq39.7$ ,\, $\tau_R=2\pi\sqrt{\vert\overline{\alpha}\vert^2+\sinh^2(r)}\simeq39.7$ respectively, is approximately equal to that at half the revival times $\frac{\tau_R}{2}$, i.e, $S_a(\tau)\simeq 0.04$ in our example. This means that at these times ($\tau_R,\,\frac{\tau_R}{2}$) the atom-field states are entangled (mixed) when an initial squeezed coherent state is considered, accordant with \cite{moya1992interaction}. In addition, as time advances, we note gradual increase in DEM and consequently the degree of mixedness since $S_a(t)$ records gradual increasing values with every increase in time;}  

\item {that the behaviour in (i) is enhanced during the JC and separately AJC interaction set at $r=1.5$ as presented in Fig.~\ref{fig:2022fent2}. The form of time evolution of $S_a(\tau)$ becomes more rapid with oscillations between [$\simeq 0.2$,\,1] characterising an increase in DEM (and so the degree of mixedness), consistent with \cite{moya1992interaction}. It is now clear that the DEM increases with an increase in \textit{r} and;}
\item{as demostrated in Fig.~\ref{fig:2022fent1} set at  $r=1$, that the evolution of $S_a(\tau)$ during the JC interaction and separately the corresponding AJC interaction are of the same form. A similar observation suffices  at $r=1.5$ as visualised in Fig.~\ref{fig:2022fent2}.} 
\end{enumerate}

\section{Conclusion}
\label{sec:conc}
We analysed separately, the interaction of a two-level atom with a single mode of an initial squeezed coherent light during the AJC,\,JC processes, and the results in the respective interactions are consistent with earlier work cited in this paper. As visualised, the nature of photon statistics and DEM take the same form during time evolution of atomic population inversion for values of squeeze parameter set in the range [1,\,1.5]. At $r>1.4$ the field system becomes dominantly super-Poissonian. We also observed that at a higher $r=1.5$ the revival peaks during atomic population inversion becomes less pronounced, irregular and ringing revivals observed at an expected collapse phase. The DEM in the respective AJC,\,JC interactions, showed that it increases with every increase in \textit{r} and interaction time $\tau=\lambda t$ albeit gradually, and so is the degree of mixedness.

\section*{Acknowledgement}

We thank Tom Mboya University and Maseno University for providing conducive environment to carry out this study.

\section*{Data availability}

All data required is available in this manuscript. 

\section*{Disclosure statement}

The authors declare no conflict of interest towards publication of this work.





\begin{thebibliography}{}
\providecommand{\url}[1]{\normalfont{#1}}
\providecommand{\urlprefix}{Available from: }

\end{thebibliography}


\begin{thebibliography}{10}
\providecommand{\url}[1]{\normalfont{#1}}
\providecommand{\urlprefix}{Available from: }

\bibitem{gerry2005introductory}
Gerry~C, Knight~PL. Introductory quantum optics. {C}ambridge {U}niversity
  {P}ress; 2005.

\bibitem{Loudon1987Squeezed}
Loudon~R, Knight~P. Squeezed light. J Mod Opt.
  1987;\hspace{0pt}34(6-7):709--759.

\bibitem{Knight2004}
Knight~PL, Bu{\v{z}}ek~V. Squeezed {S}tates: {B}asic {P}rinciples. In:
  Drummond~PD, Ficek~Z, editors. Quantum {S}queezing. Berlin, {H}eidelberg:
  Springer {B}erlin {H}eidelberg; 2004. p. 3--32.

\bibitem{zaheer1990advances}
Zaheer~K, Zubairy~M. In: Bates~D, Bederson~B, editors. Advances in {A}tomic,
  {M}olecular and {O}ptical {P}hysics. Vol.~28. New {Y}ork: {A}cademic {P}ress;
  1987. p. 143.

\bibitem{satyanarayana1989ringing}
Satyanarayana~MV, Rice~P, Vyas~R, et~al. Ringing revivals in the interaction of
  a two-level atom with squeezed light. JOSA B. 1989;\hspace{0pt}6(2):228--237.

\bibitem{schleich1987oscillations}
Schleich~W, Wheeler~J. Oscillations in photon distribution of squeezed states.
  JOSA B. 1987;\hspace{0pt}4(10):1715--1722.

\bibitem{schleich1988area}
Schleich~W, Walls~D, Wheeler~J. Area of overlap and interference in phase space
  versus {W}igner pseudoprobabilities. {P}hys {R}ev A.
  1988;\hspace{0pt}38(3):1177.

\bibitem{teich1989squeezed}
Teich~MC, Saleh~BE. Squeezed state of light. Quant Opt.
  1989;\hspace{0pt}1(2):153.

\bibitem{omolo2021anti}
Omolo~JA. The anti-{J}aynes-{C}ummings model is solvable: quantum {R}abi model
  in rotating and counter-rotating frames; following the experiments. arXiv
  preprint arXiv:210309546. 2021;\hspace{0pt}.

\bibitem{jaynes1963comparison}
Jaynes~ET, Cummings~FW. Comparison of quantum and semiclassical radiation
  theories with application to the beam maser. Proceedings of the IEEE.
  1963;\hspace{0pt}51(1):89--109.

\bibitem{born1999principles}
Born~M, Wolf~E. Principles of optics: electromagnetic theory of propagation,
  interference and diffraction of light, 7th edition. {C}ambridge {U}niversity
  {P}ress; 1999.

\bibitem{scully1997quantum}
Scully~MO, Zubairy~MS, et~al. Quantum optics. Cambridge University Press; 1997.

\bibitem{mandel1979sub}
Mandel~L. Sub-{P}oissonian photon statistics in resonance fluorescence. Opt
  Lett. 1979;\hspace{0pt}4(7):205--207.

\bibitem{teich1988photon}
Teich~MC, Saleh~BE. I photon bunching and antibunching. In: Progress in optics.
  Vol.~26. Elsevier; 1988. p. 1--104.

\bibitem{mandel1995optical}
Mandel~L, Wolf~E. Optical coherence and quantum optics. Cambridge {U}niversity
  {P}ress; 1995.

\bibitem{moya1992interaction}
Moya-Cessa~H, Vidiella-Barranco~A. Interaction of squeezed light with two-level
  atoms. J Mod Opt. 1992;\hspace{0pt}39(12):2481--2499.

\bibitem{eberly1980}
Eberly~J, Narozhny~N, Sanchez-Mondragon~J. {P}eriodic {S}pontaneous {C}ollapse
  and {R}evival in a {S}imple {Q}uantum {M}odel. {P}hys {R}ev {L}ett.
  1980;\hspace{0pt}44(20):1323--1325.

\end{thebibliography}
\end{document}